\documentclass[aps,prd,onecolumn,groupedaddress,showpacs,nofootinbib,amssymb]{revtex4-2}
\usepackage[dvips]{graphicx}
\usepackage{amssymb}
\usepackage{amsmath}
\usepackage{graphicx,,color}
\usepackage{amsfonts}
\usepackage{bm}
\usepackage{cancel}
\usepackage{comment}

\newcommand\be{\begin{equation}}
\newcommand\ee{\end{equation}}

\newcommand\e{\mathrm{e}}

\allowdisplaybreaks[4]

\begin{document}

\tolerance=5000

\title{Maximal Masses of White Dwarfs for Polytropes in $R^2$ Gravity and Theoretical Constraints}
\author{A.V. Astashenok$^{1}$}
\email{aastashenok@kantiana.ru}
\author{S.~D.~Odintsov,$^{2,3}$}
\email{odintsov@ice.cat}
\author{V.K. Oikonomou,$^{4}$}
\email{voikonomou@gapps.auth.gr,v.k.oikonomou1979@gmail.com}
\affiliation{$^{1)}$ Institute of Physics, Mathematics and IT, I.
Kant Baltic Federal University, Kaliningrad, 236041 Russia,\\
$^{2)}$ ICREA, Passeig Luis Companys, 23, 08010 Barcelona, Spain\\
$^{3)}$ Institute of Space Sciences (ICE,CSIC) C. Can Magrans s/n,
08193 Barcelona, Spain\\
$^{4)}$ Department of Physics, Aristotle University of
Thessaloniki, Thessaloniki 54124, Greece }

\date{\today}
\tolerance=5000

\begin{abstract}
We examine the Chandrasekhar limit for white dwarfs in $f(R)$
gravity, with a simple polytropic equation of state describing
stellar matter. We use the most popular $f(R)$ gravity model,
namely the $f(R)=R+\alpha R^2$ gravity, and calculate the
parameters of the stellar configurations with polytropic equation
of state of the form $p=K\rho^{1+1/n}$ for various values of the
parameter $n$. In order to simplify our analysis we use the
equivalent Einstein frame form of $R^2$-gravity which is basically
a scalar-tensor theory with well-known potential for the scalar
field. In this description one can use simple approximations for
the scalar field $\phi$ leaving only the potential term for it.
Our analysis indicates that for the non-relativistic case with
$n=3/2$, discrepancies between the $R^2$-gravity and General
Relativity can appear only when the parameter $\alpha$ of the
$R^2$ term, takes values close to maximal limit derived from the
binary pulsar data namely $\alpha_{max}=5\times 10^{15}$ cm$^2$.
Thus, the study of low-mass white dwarfs can hardly give
restrictions on the parameter $\alpha$. For relativistic
polytropes with $n=3$ we found that Chandrasekhar limit can in
principle change for smaller $\alpha$ values. The main conclusion
from our calculations is the existence of white dwarfs with large
masses $\sim 1.33 M_{\odot}$, which can impose more strict limits
on the parameter $\alpha$ for the $R^2$ gravity model.
Specifically, our estimations on the parameter $\alpha$ of the
$R^2$ model is $\alpha \sim 10^{13}$ cm$^2$.
\end{abstract}


\maketitle

\section{Introduction}

The accelerated expansion of the Universe \cite{Perlmutter,
Riess1, Riess2} has been confirmed by various observations from
moment of its discovering in 1998. Apart from data based on
standard candles, these observations include of data for cosmic
microwave background anisotropy \cite{Spergel}, shear due to the
gravitational weak lensing \cite{Schmidt}, data about absorption
lines in Lyman-a-forest \cite{McDonald} and others. Now the
question is what causes the late-time cosmological acceleration.
To date, there no clear answer exists on this issue,  and this
question remains one of the puzzles in cosmology and theoretical
physics. One solution for this late-time acceleration problem is
based on the assumption that the Universe contains some
non-standard cosmic fluid with negative pressure. This fluid ({\it
dark energy}) is distributed in the Universe homogeneously. In the
simplest approach, the dark energy is nothing else than the vacuum
energy (or cosmological constant). For a satisfactory agreement
with observational data density of vacuum energy should be nearly
72\% of the global energy budget of the Universe. Only 4 \% of
Universe energy consists of baryonic matter. The remaining 24\% is
cold dark matter (CDM) which is also mysterious in its nature.
Various unknown hypothetical particles could constitute this
``dark sector'' of the Universe, for example weakly interacting
massive particles or/and axions or axion-like particles.

To date, the most successful model for the late-time evolution of
the Universe is the $\Lambda$-Cold-Dark-Matter ($\Lambda$CDM)
model. Although standard cosmology describes observational data
with high precision, it has several shortcomings, from a
theoretical viewpoint. If the cosmological constant constitutes
the ``vacuum state'' of the gravitational field, one need have to
explain the very large discrepancy of the magnitude between its
observed value at a cosmological level and the one predicted by
any quantum field theories \cite{Weinberg}. This discrepancy is
also known as the cosmological constant problem, which is one of
the most fundamental problems of the $\Lambda$CDM model. Apart
from this theoretical problem, if one sticks with the standard
General Relativity (GR) description of late-time and uses scalar
fields to describe the late-time evolution, the slightly phantom
nature allowed by the observational data, makes the GR description
unappealing, since tachyons are needed in order to successfully
describe the slightly phantom evolution. Another appealing
theoretical description of the dark energy era is offered by
modified gravity in its various forms
\cite{reviews1,reviews2,reviews4,book,reviews5,reviews6,dimo},
which extend GR directly. Modified gravity theories have the
advantage of describing successfully and in a minimal way the
late-time era without the use of phantom scalars, and in many
cases, the unification of the inflationary era with the early-time
acceleration is achieved. However, when considering modifications
of GR, a holistic approach compels to consider not only possible
manifestations of such theories at a cosmological level, but also
at an relativistic astrophysical level, also because strong
gravitational regimes could be considered if GR is the weak field
limit of some more complicated effective gravitational theory.

Usually in this context, neutron stars (NSs) serve as the perfect
candidates for studying modified gravity effects at an
astrophysical level, and in the literature there exist various
works in this research line, see for example,
\cite{Astashenok:2020qds,Astashenok:2021peo,Capozziello:2015yza,Astashenok:2014nua,Astashenok:2014pua,Astashenok:2013vza,Arapoglu:2010rz,Panotopoulos:2021sbf,Lobato:2020fxt,Oikonomou:2021iid,Odintsov:2021nqa,Odintsov:2021qbq,Katsuragawa},
and for a recent review see \cite{Wojnar3}. For a simple $R^2$
gravity, the calculations indicate that the effective
gravitational mass of NSs increases although the value of such
increase is not large. The scalar curvature $R$ outside the NS
doesn't drop to zero as for for the Schwarzschild solution in GR,
but asymptotically approaches zero at the spatial (and from a
calculational point of view at the numerical) infinity. The
gravitational mass parameter at the star surface decreases in
comparison with GR for same density of nuclear matter at the
center of star, but contributions to the gravitational mass are
obtained from regions beyond the surface of the star, at which
$R\neq 0$, thus the net result is a total increase of the
gravitational mass. This effect in NSs may explain in an appealing
way the hyperon problem and having soft equations of state with
large NSs masses, beyond the stretch of GR limits.

This interesting result can provide a clear cut description for
large mass NSs
\cite{Pani:2014jra,Doneva:2013qva,Horbatsch:2015bua,Silva:2014fca,Chew:2019lsa,Blazquez-Salcedo:2020ibb,Motahar:2017blm,Oikonomou:2021iid,Odintsov:2021nqa,Odintsov:2021qbq}.
In light of the relatively recent GW190814 event, modified gravity
serves as a cutting edge probable description of nature in limits
where GR needs to be supplemented by a Occam's razor compatible
theory. Indeed, solutions such as strange stars, rely on QCD,
which directly changes the physics of hydrodynamic equilibrium of
compact stellar objects. Modified gravity on the other hand does
not change the way of thinking for relativistic objects, just
changes the gravitational theory which controls the hydrodynamic
equilibrium.

From the above line of reasoning, in this work we consider another
class of compact objects, namely white dwarfs. White dwarfs are usually the final state of evolution for stars with masses up to 8 - 10.5 $M_\odot$ \cite{Shapiro}. 
After the hydrogen-fusing period of a main-sequence star ends, star will expand to a red giant. Due to the $\alpha$-process helium fuses to carbon and oxygen. For low and medium star masses core temperature is insufficient to fuse carbon and after shedding of outer layers remnant composed of carbon and oxygen forms. For main-sequence stars with more large masses (in range of 8-10.5 $M_\odot$, the core temperature will exceed $10^{9}$ K and carbon fuses but not neon. In this case oxygen–neon–magnesium white dwarf remains \cite{Werner}. Also helium white dwarfs can forms due to mass loss in binary systems \cite{Liebert}.

Electron degeneracy pressure supports a white dwarf in equilibrium. One of the consequence of this is the existence of a limit for mass of white dwarf which cannot be exceeded without collapsing to a neutron star This value is known as Chandrasekhar limit \cite{Chandrasekhar}. For a non-rotating white dwarf maximal mass is approximately $5.7M_{\odot}/\mu_{e}^{2}$, where $\mu_e$ is the average molecular weight per electron of the star. It is interesting to consider question about Chandrasekhar limit in modified gravity theories.

The authors of Ref.
\cite{Kalita} considered the model $f(R)=R+\alpha R^2 (1+\gamma
R)$ and obtained for very large range of values of the parameters
$\alpha$ and $\gamma$ ($\gamma = 4\times 10^{16}$ cm$^2$) a
considerable increase of the Chandrasekhar limit for white dwarfs
(up to $2.95 M_{\odot}$). This result is interesting but the
existence of such white dwarfs is not confirmed by observational
data. More realistic models of white dwarfs in Palatini $f(R)$
gravity (in such approach there is no extra degree of freedom for
the gravitational sector) are investigated in Refs.
\cite{Wojnar1,Wojnar2}. Although the central densities of white
dwarfs are not so large as in the cores of NSs, one can expect
that effects of modified gravity would take place due to the
larger radii of white dwarfs (in principle one can say about
``cumulative effects''). Our calculations confirm this conclusion.
We start off with the usual equations describing non-rotating star
in equilibrium and we use the Lane-Emden equation for matter with
a polytropic equation of state. In GR one already needs to
consider the Tolman-Oppenheimer-Volkoff (TOV) equations. For white
dwarfs, the relativistic effects are negligible of course, but it
is interesting to compare these corrections with possible
influences of modified gravity. This is the main reason for which
we consider these equations. The next step is to investigate
solutions for equations describing star configuration in $R^2$
gravity. Besides the mass parameter, the density and pressure in
$f(R)$ gravity, another independent variable arises, namely the
scalar curvature. In GR for non-rotating stars, the scalar
curvature is defined by trace of energy-momentum tensor. But in
$f(R)$ gravity the scalar curvature should be also considered when
studying the dynamical equilibrium.

We consider simple polytropic equations of state for matter, of
the form, $p=K\rho^{1+1/n}$ describing the white dwarfs. For
non-relativistic electrons, the index of the polytrope is $n=3/2$,
and for relativistic ones, the index is $n=3$. For our analysis it
is useful to consider the equivalent Brans-Dicke theory with a
scalar field. Assuming that the derivatives of the scalar field
are negligible, in comparison with potential term, one obtains
simple equations which are similar to the TOV equations, but with
additional terms for the pressure and the density. The net effect
of this scalar field is that it reduces the gravitational mass and
simultaneously increases the radius of the stellar configuration.

\section{The Lane-Emden Equation and the TOV Equations}

The hydrodynamic equilibrium of non-rotating stars in Newtonian
gravity is described by the following well-known equations (hereinafter we use
Geometrized units in which $c=G=1$):
\begin{equation}\label{eq1}
    \frac{dm}{dr}=4\pi r^2 \rho,
\end{equation}
\begin{equation}\label{eq2}
    \frac{dp}{dr}=-\frac{m\rho}{r^2},
\end{equation}
where $\rho$ and $p$ are the density and the pressure of stellar
matter respectively. For white dwarfs, one can use polytropic
equation of state of the form,
\begin{equation}
    p=K\rho^{1+1/n}\, ,
\end{equation}
where $K$, $n$ are constants. We assume that the speed of light is
$c=1$ and therefore pressure and density have the same dimensions.
In this case, it is useful to define the following dimensionless
functions $\theta$ and $\mu$ and the coordinate variable $x$:
$$
\rho = \rho_{c} \theta^{n},\quad m = \mu\rho_c a^3, \quad r=a x,
$$
$$
a = \left(\frac{(n+1)K\rho_{c}^{1/n-1}}{4\pi}\right)^{1/2}.
$$
In terms of dimensionless variables, Eqs. (\ref{eq1}), (\ref{eq2})
are rewritten in the following way
\begin{equation}\label{eq1-2}
    \frac{d\mu}{dx} = 4\pi x^2 \theta^{n},
\end{equation}
\begin{equation}\label{eq2-2}
    \frac{d\theta}{dx} = -\frac{\mu}{4\pi x^2}.
\end{equation}
These equations are equivalent to well-known Lane-Emden equation,
\begin{equation}
    \frac{1}{x^2}\frac{d}{dx}\left(x^2\frac{d\theta}{dx}\right)=-\theta^n.
\end{equation}
For more exact description one need to account relativistic theory
of gravity, in which case one needs to replace Eqs. (\ref{eq1})
and (\ref{eq2}) by the TOV equations,
\begin{equation}\label{OV-1}
    \frac{dm}{dr} = 4\pi \rho r^2,
\end{equation}
\begin{equation}\label{OV-2}
    \frac{dp}{dr} = - (\rho + p) \frac{m+4\pi p r^3}{r^2\left(1-\frac{2m}{r}\right)}.
\end{equation}
The second equation for the polytropic EoS and variables $\mu$,
$\theta$ can be reduced to,
\begin{equation}\label{OV-3}
    \frac{1}{1+4\pi\beta\theta/(n+1)}\frac{d\theta}{dx} = -\frac{1}{4\pi} \frac{\mu+16\pi^2 (n+1)^{-1} x^{3}\beta\theta^{n+1}}{x(x-2\beta \mu)}.
\end{equation}
The equation for the gravitational parameter $\mu$ remains the
same. Here $\beta$ is a dimensionless small parameter,
$$
\beta = \rho_c a^2 = \frac{n+1}{4\pi}K\rho_{c}^{1/n}<<1\, ,
$$
for typical densities at the center of white dwarfs. Up to
densities $10^6$ g/cm$^3$ one can use polytropic equation with
$n=3/2$ and,
$$
K=1.0036\times 10^{13}/\mu_{e}^{5/3}\, ,
$$
in CGS-system. Here $\mu_e$ is average molecular weight per one
electron. We considered $\mu_e=2$. One can solve the equations
(\ref{eq2-2}) and (\ref{OV-3}) numerically and obtain physical
parameters of polytropes, that is the mass and radius.

The integration of the Lane-Emden equation for $n=3/2$ gives that
for $x=x_{f}=3.653$ $\theta(x_f)=0$ and $\mu(x_{f})=34.106$. These
results change in GR, but are quantitatively negligible. In Fig. 1
we present the corresponding relation between the mass, radius and
the central density of the stellar configurations for central
densities between $10^5$ and $10^6$ $\mbox{g/cm}^3$.
\begin{figure}
    \centering
    \includegraphics[scale=0.4]{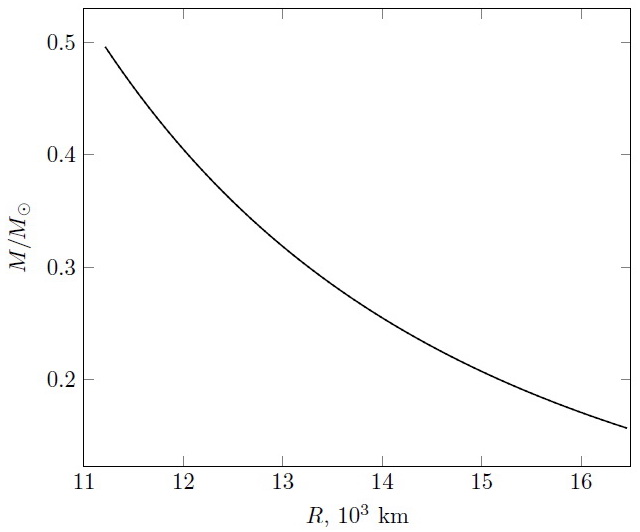}\includegraphics[scale=0.4]{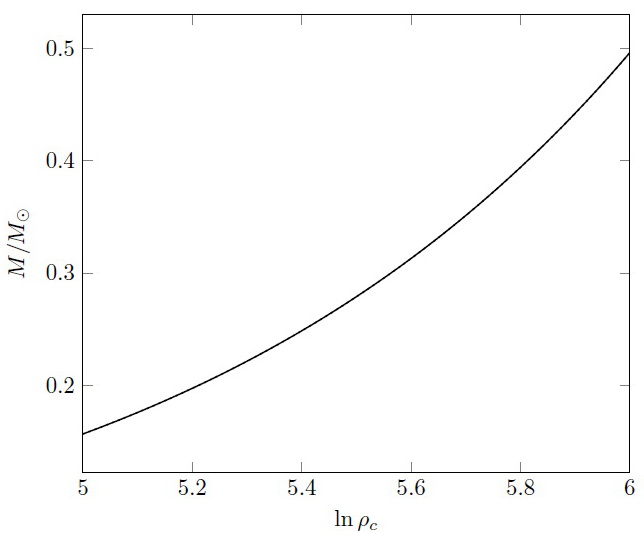}\\
    \caption{Mass-radius and mass-central density diagrams for stellar configurations with $n=3/2$.}
    \label{fig:1}
\end{figure}
 \begin{figure}
    \centering
    \includegraphics[scale=0.4]{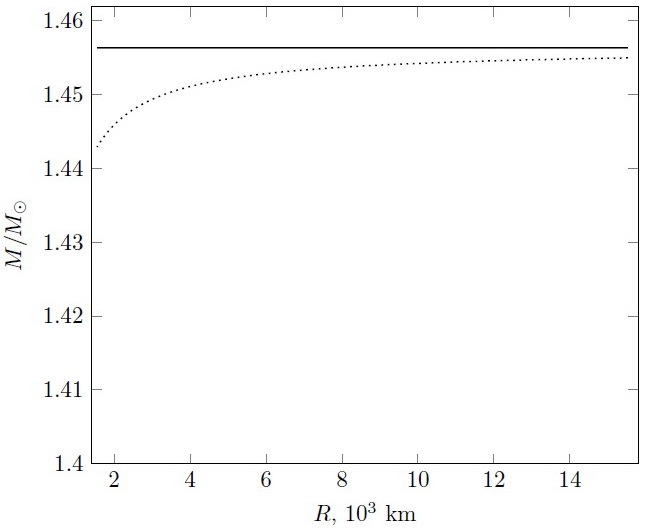}\includegraphics[scale=0.4]{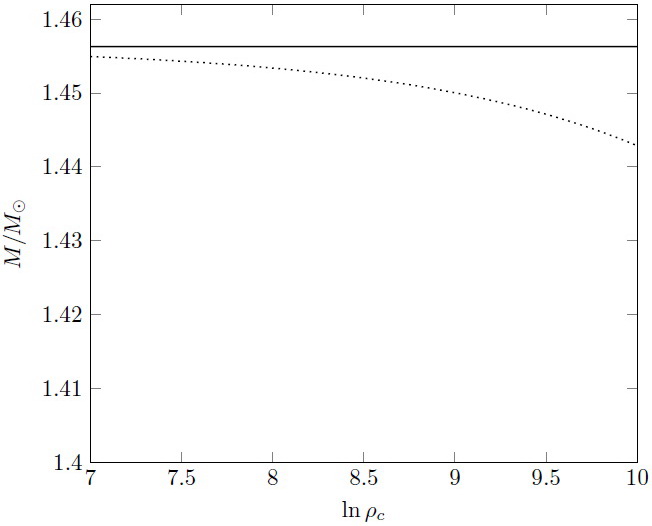}\\
    \caption{Mass-radius and mass-central density diagrams for polytropic
stellar configurations with index $n=3$. The solid and dotted
lines correspond to solution of Lane-Emden equation and
relativistic Eqs. (\ref{eq1-2}), (\ref{OV-3}) correspondingly.}
    \label{fig:2}
\end{figure}
For relativistic densities ($\rho>>10^6$ g/cm$^3$), stellar matter
in white dwarfs can be described by polytropic equations with
$n=3$. The parameter $K$ in this case is,
$$
K=1.2435\times 10^{15}/\mu_{e}^{4/3} \quad \mbox{CGS}.
$$
For the case $n=3$, the mass of star does not depend on the
central density and $\mu(x_{f})=25.362$ for $x_f=6.896$ in
Geometrized units. This corresponds to the Chandrasekhar limit
$M=1.456 M_{\odot}$. The relativistic effects again slightly
change this result (see Fig. 2). Now we investigate how these
results change in modified gravity.

\section{Spherically symmetric stellar configurations in f(R)-gravity}

For $f(R)$ gravity the standard Einstein-Hilbert action with
scalar curvature $R$ is replaced by the following:
\begin{equation}\label{action}
S=\frac{1}{16\pi}\int d^4x \sqrt{-g}f(R) + S_{{\rm matter}},
\end{equation}
where $g$ is determinant of the metric $g_{\mu\nu}$ and $S_{\rm
matter}$ is the action of the standard perfect fluid matter.

The spherically symmetric spacetime that describes the
non-rotating star has the following form,
\begin{equation}\label{metric}
    ds^2= -e^{2\psi}dt^2 +e^{2\lambda}dr^2 +r^2 d\Omega^2.
\end{equation}
Here $\psi$ and $\lambda$ are two independent functions of the
radial coordinate. Varying the action with respect to metric
tensor elements gives the Einstein equation in $f(R)$ gravity:
\begin{equation}\label{field}
f'(R)G_{\mu \nu }-\frac{1}{2}(f(R)-f'(R)R)g_{\mu \nu }-(\nabla
_{\mu }\nabla _{\nu }-g_{\mu \nu }\Box )f'(R)=8 \pi T_{\mu \nu }.
\end{equation}
Here $G_{\mu\nu}=R_{\mu\nu}-\frac{1}{2}Rg_{\mu\nu}$ is the
Einstein tensor, comma means derivative on argument of function
$f(R)$ and $T_{\mu \nu }$ is the energy--momentum tensor. For
perfect fluid components of $T_{\mu\nu}$ is
$T_{\mu\nu}\mbox{diag}(e^{2\psi}\rho, e^{2\lambda}p, r^2p,
r^{2}\sin^{2}\theta p)$. The components of (\ref{field}) yield the
TOV equations in frames of $f(R)$ gravity, and explicitly have the
following form: \be \label{TOV1}
\frac{f'(R)}{r^2}\frac{d}{dr}\left [r\left(1-e^{-2\lambda
}\right)\right]=8\pi
\rho+\frac{1}{2}\left(f'(R)R-f(R)\right)+e^{-2\lambda}\left[\left(\frac{2}{r}-\frac{d\lambda}{dr}\right)\frac{d
f'(R)}{dr}+\frac{d^{2}f'(R)}{dr^2}\right] \ee

\be \label{TOV2} \frac{f'(R)}{r}
\left[2e^{-2\lambda}\frac{d\psi}{dr}-\frac{1}{r}\left(1-e^{-2\lambda}\right)\right]
=8\pi
p+\frac{1}{2}\left(f'(R)R-f(R)\right)+e^{-2\lambda}\left(\frac{2}{r}+\frac{d\psi}{dr}\right)\frac{df'(R)}{dr}\ee
The third equation can be obtained from the conservation law of
the energy-momentum tensor
$$
\nabla^\mu T_{\mu\nu}=0
$$
and gives,
\begin{equation}\label{hydro}
    \frac{dp}{dr}=-(\rho
    +p)\frac{d\psi}{dr}.
\end{equation}
The equation for the scalar curvature $R$ can be obtained by
taking of the trace of (\ref{field}). We have for $R=R(r)$ the
following equation:
\begin{equation}\label{TOV3}
3\triangle_{r} (f(R)-R)+f'(R)R-2f(R)=-{8\pi}(\rho-3p),
\end{equation}
where $\triangle_{r}$ is radial part of the 3-dimensional Laplace
operator for metric (\ref{metric}),
$$
e^{2\lambda}\triangle_{r}=\frac{d^2}{dr^2}+\left(\frac{2}{r}+\frac{d\psi}{dr}-\frac{d\lambda}{dr}\right)\frac{d}{dr}.
$$

One needs to solve Eqs. (\ref{TOV1}), (\ref{TOV2}) outside the
star with $\rho=p=0$. At the surface of star ($r=r_{s}$, $\rho=p=0$), the
junction conditions should be satisfied,
$$\lambda_{in}(r_{s})=\lambda_{out}(r_{s}),\quad
R_{in}(r_{s})=R_{out}(r_{s}), \quad
R'_{in}(r_{s})=R'_{out}(r_{s}).$$ 
The gravitational mass parameter
$m(r)$ is defined from $\lambda$ through the following relation,
\begin{equation}
\label{mass}
    e^{-2\lambda}=1-\frac{2m}{r}.
\end{equation}
The asymptotic flatness requirement gives the constraint on scalar
curvature and mass parameter,
$$\lim_{r\rightarrow\infty}R(r)=0,
\lim_{r\rightarrow\infty}m(r)=\mbox{const}.$$

\section{Equivalent Scalar-Tensor Theory}

It is useful to consider the description of the $f(R)$ gravity
stellar configuration in terms of the corresponding scalar-tensor
theory. For such a theory, the difference between the resulting
equations and (\ref{OV-1}), (\ref{OV-2}) for polytropes is more
transparent. In the end for the construction of the $M-R$ graphs
we shall use the Jordan frame transformed expressions for the mass
and radii. For $f(R)$ gravity, the equivalent Brans-Dicke theory with scalar field $\Phi$
has the following action,
 \be S_{g}=\frac{1}{16\pi}\int
d^{4} x \sqrt{-g}\left(\Phi R - U(\Phi)\right). \ee Here the
scalar field is $\Phi=f'(R)$ and the potential is
$U(\Phi)=Rf'(R)-f(R)$. The transformation $\tilde{g}_{\mu\nu}=\Phi
g_{\mu\nu}$ for coordinates allows to write the action in the
Einstein frame \be S_{g}=\frac{1}{16\pi}\int d^{4} x
\sqrt{-\tilde{g}}\left(\tilde{R}
-2\tilde{g}^{\mu\nu}\partial_{\mu}\phi\partial_{\nu}\phi-4V(\phi)\right),
\ee where $\phi=\sqrt{3}\ln \Phi/2$ and the redefined potential in
the Einstein frame $V(\phi)$ is
$V(\phi)=\Phi^{-2}(\phi)U(\Phi(\phi))/4$. For redefined spacetime
metrics, we take the expression which formally coincides with
(\ref{metric}), but with different functions $\tilde{\psi}$ and
$\tilde{\lambda}$, that is, \be \label{metric2}
d\tilde{s}^{2}=\Phi
ds^{2}=-e^{2\tilde{\psi}}dt^{2}+e^{2\tilde{\phi}}\tilde{dr}^{2}+\tilde{r}^2d\Omega^2.
\ee In Eq. (\ref{metric2}) we have $\tilde{r}^2=\Phi r^{2}$,
$e^{2\tilde{\psi}}=\Phi e^{2\psi}$ and from the equality
$$
\Phi e^{2\lambda}dr^{2}=e^{2\tilde{\lambda}}d\tilde{r}^{2}
$$
it follows that,
$$
e^{-2\lambda}=e^{-2\tilde{\lambda}}\left(1-\tilde{r}\phi'(\tilde{r})/\sqrt{3}\right)^{2}.
$$
Therefore, the mass parameter $m(r)$ can be obtained from
$\tilde{m}(\tilde{r})$ as \be m(\tilde{r})=
\frac{\tilde{r}}{2}\left(1-\left(1-\frac{2\tilde{m}}{\tilde{r}}\right)\left(1-\tilde{r}\phi'(\tilde{r})/\sqrt{3}\right)^{2}\right)e^{-\phi/\sqrt{3}}
\ee The resulting equations for the metric functions
$\tilde{\lambda}$ and $\tilde{\psi}$ is very similar to the TOV
equations with redefined energy and pressure, and with additional
terms with the energy density and pressure of the scalar field
$\phi$ being: \be
\label{TOV1-1} \frac{1}{\tilde{r}^2}\frac{d \tilde{m}}{d\tilde{r}}=4\pi
e^{-4\phi/\sqrt{3}}\rho+\frac{1}{2}\left(1-\frac{2\tilde{m}}{\tilde{r}}\right)\left(\frac{d\phi}{d\tilde{r}}\right)^{2}+V(\phi),
\ee

\be \label{TOV2-1} \frac{1}{p+\rho}\frac{dp}{d\tilde{r}}=-\frac{\tilde{m} + 4\pi
e^{-4\phi/\sqrt{3}} p \tilde{r}^3}{\tilde{r}(\tilde{r}-2\tilde{m})}-
\frac{\tilde{r}}{2}\left(\frac{d\phi}{d\tilde{r}}\right)^{2}+\frac{\tilde{r}^2
V(\phi)}{\tilde{r}-2\tilde{m}}+\frac{1}{\sqrt{3}}\frac{d\phi}{d\tilde{r}} ,\ee The hydrostatic equilibrium condition is
rewritten as,
\begin{equation}\label{hydro-1}
    \frac{dp}{d\tilde{r}}=-(\rho
    +p)\left(\frac{d\psi}{d\tilde{r}}-\frac{1}{\sqrt{3}}\frac{d\phi}{d\tilde{r}}\right).
\end{equation}
Finally the last equation of motion for scalar field is equivalent
to Eq. (\ref{TOV3}) in $f(R)$ theory: \be \label{TOV3-1}
\triangle_{\tilde{r}} \phi-\frac{dV(\phi)}{d\phi}=-\frac{4\pi}{\sqrt{3}}
e^{-4\phi/\sqrt{3}}(\rho-3p). \ee The potential $V(\phi)$ can be
written in explicit form only for simple $f(R)$ models. For
example for $f(R)=R+\alpha R^2$ one can obtain that, \be
V(\phi)=\frac{1}{16\alpha}\left(1-e^{-2\phi/\sqrt{3}}\right)^2.
\ee For simple power-law models of the form $f(R)=R+\alpha R^{m}$
we have, \be
V(\Phi)=D\Phi^{-2}\left(\Phi-1\right)^{\frac{m}{m-1}}, \quad
D=\frac{m-1}{m^{\frac{m}{m-1}}}\alpha^{\frac{1}{1-m}}, \quad
\Phi=e^{2\phi/\sqrt{3}}. \ee Passing to dimensionless variables
$\tilde{\mu}$ and $\theta$ we obtain the following equations,
\begin{equation}\label{TOVmod-1}
    \frac{d\tilde{\mu}}{dx} = 4\pi \tilde{x}^2 \theta^{n} \e^{-4\phi/\sqrt{3}} + \frac{\tilde{x}^2}{\beta}\left(\frac{1}{2}\left(1-\frac{2\beta \tilde{\mu}}{\tilde{x}}\right)\left(\frac{d\phi}{d\tilde{x}}\right)^{2}+v(\phi)\right),
\end{equation}
\begin{equation}\label{TOVmod-2}
    \frac{1}{1+4\pi\beta\theta/(n+1)}\frac{d\theta}{d\tilde{x}}=-\frac{1}{4\pi}\frac{\tilde{\mu}+16\pi^{2}\beta\theta^{n+1}\tilde{x}^3 e^{-4\phi/\sqrt{3}}}{\tilde{x}(\tilde{x}-2\tilde{\beta}\tilde{\mu})}-
\end{equation}
$$
-\frac{\tilde{x}^2}{4\pi \beta(\tilde{x}-2\beta\tilde{\mu})}\left(\frac{1}{2}\left(1-\frac{2\beta \tilde{\mu}}{\tilde{x}}\right)\left(\frac{d\phi}{d\tilde{x}}\right)^{2}-v(\phi)\right)+\frac{1}{4\sqrt{3}\pi\beta}\frac{d\phi}{d\tilde{x}}
$$
Here we introduced the dimensionless potential $v(\phi)$,
$$
v(\phi)=a^2 V(\phi).
$$
The equation for scalar field $\phi$ after some calculations can
be written in the following form,
\begin{equation}\label{EqSc}
    \left(1-\frac{2\beta \tilde{\mu}}{\tilde{x}}\right)\left(\frac{d^2 \phi}{d\tilde{x}^2}+\left(\frac{2}{\tilde{x}}-\frac{4\pi\beta}{1+4\pi\beta/(n+1)}\frac{d\theta}{d\tilde{x}}+\frac{1}{\sqrt{3}}\frac{d\phi}{d\tilde{x}}\right)\frac{d\phi}{d\tilde{x}}\right)+
\end{equation}
$$
+\left(\frac{\beta \tilde{\mu}}{\tilde{x}^2}-\frac{\beta}{\tilde{x}}\frac{d\tilde{\mu}}{d\tilde{x}}\right)\frac{d\phi}{d\tilde{x}}-\frac{dv}{d\phi} = -\frac{4\pi\beta}{\sqrt{3}}e^{-4\phi/\sqrt{3}}\theta^{n}\left(1-12\pi\beta\theta/(n+1)\right).
$$
Equations (\ref{TOVmod-1}), (\ref{TOVmod-2}) with (\ref{EqSc}) can
be solved numerically for various values of the parameter $n$. One
needs to impose following conditions at $\tilde{x}=0$ for the numerical
integration,
$$
\theta(0)=1,\quad \tilde{\mu}(0)=0, \quad \phi(0)=\phi_{0}, \quad \frac{d\phi}{d\tilde{x}}=0.
$$
The condition for $\phi_{0}$ should correspond to a solution with
flat asymptotic behavior at spatial infinity (numerical infinity),
that is,
$$\phi\rightarrow 0 \quad \mbox{at} \quad x\rightarrow \infty. $$

{It is convenient to analyse system of equations in Einstein frame and then after calculations go back to Jordan frame.}

\section{Perturbative Analysis of the Solution}

For $R^2$ gravity, the potential in Geometrized units is,
\begin{equation}\label{simple}
    v(\phi)=\frac{1}{16\tilde{\alpha}}\left(1-e^{-2\phi/\sqrt{3}}\right)^{2},\quad \tilde{\alpha}=\alpha/a^2.
\end{equation}
The authors of Ref. \cite{Naf} estimated the upper limit for
$\alpha$ as $\sim 5\times 10^{15}$ cm$^2$  from binary pulsar
data. For $n=3/2$ scale $a$ varies from $3\times 10^8$ cm to
$4.5\times 10^8$ cm (for central densities $10^5<\rho_c<10^6$
g/cm$^3$. The value of scalar field is very small and therefore
one can use Taylor expansion for the potential leaving only the
first non-zero term:
$$
v(\phi)=\frac{\phi^2}{12\tilde{\alpha}}.
$$
Since the parameter $\alpha/a^2$ is negligible, one can expect
that the potential term $v(\phi)$ is very large in comparison with
kinetic term $(d\phi/d\tilde{x})^{2}$ and we may omit these terms in the
right hand side of Eqs. (\ref{TOVmod-1}), (\ref{TOVmod-2}).
Considering Eq. (\ref{EqSc}) in this context one can neglect
derivatives of scalar field in comparison with term $dv/d\phi\sim
\phi$. In the right hand side of Eq. (\ref{EqSc}) we leave only
zero term for exponent (since $\beta$ is already a small
parameter) and in effect we obtain the following relation for the
scalar field,
\begin{equation}\label{approx}
    \frac{\phi}{6\tilde{\alpha}}=\frac{4\pi\beta}{\sqrt{3}}\theta^{n}\left(1-12\pi\beta\theta/(n+1)\right).
\end{equation}
Keeping in Eq. (\ref{TOVmod-2}) the last term with first
derivative for scalar field and potential terms which $\sim
\tilde{\alpha}$, we will integrate equations Eqs.
(\ref{TOVmod-1}), (\ref{TOVmod-2}) for polytropic EoSs with
$n=3/2$ and $n=3$ and we shall compare the results with Newtonian
gravity and General Relativity. If the approximation
(\ref{approx}) is valid, then the scalar field at the surface of
the star is zero and the gravitational mass parameter $\tilde{m}$
coincides with $m$ at the star surface $r=\tilde{r}=R_{s}$ where
$R_s$ is the radius of star.
\begin{figure}
    \centering
    \includegraphics[scale=0.4]{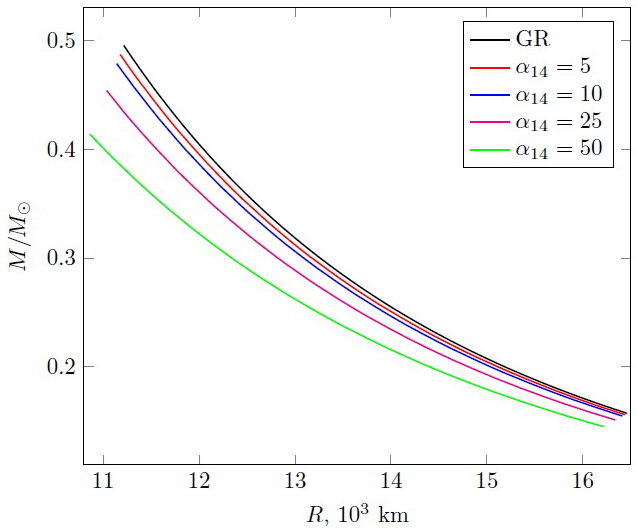}\includegraphics[scale=0.4]{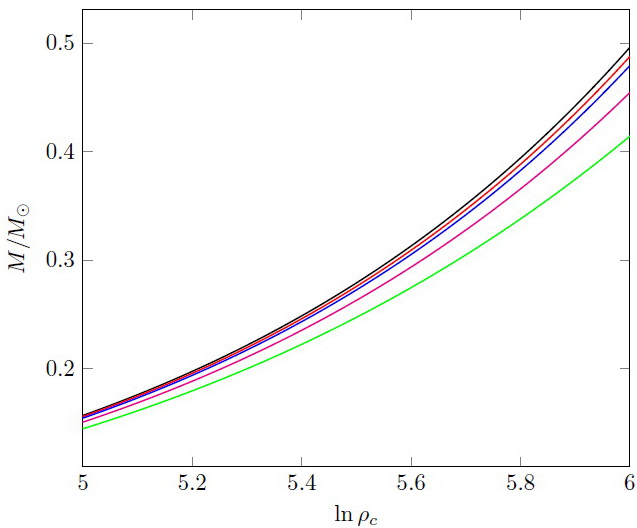}\\
    \caption{Mass-radius and mass-central density diagrams for stellar
configurations with $n=3/2$ for various $\alpha$ (symbol
$\alpha_{n}$ hereinafter means that $\alpha$ is given in units of
$10^{n}$ cm$^2$) in comparison with General Relativity.}
    \label{fig:3}
\end{figure}
\begin{figure}
    \centering
    \includegraphics[scale=0.4]{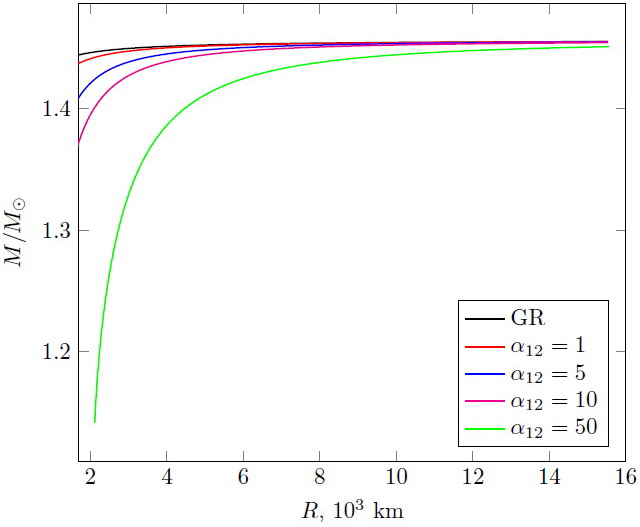}\includegraphics[scale=0.4]{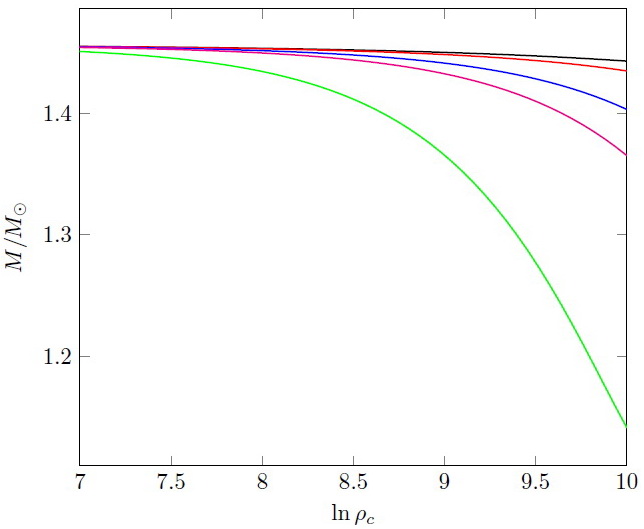}\\
    \caption{Mass-radius and mass-central density diagrams for stellar
configurations with $n=3$ in scalar-tensor gravity for various
$\alpha$ in comparison with General Relativity.}
    \label{fig:4}
\end{figure}

\begin{table}[ht]
\begin{center}
\begin{tabular}{|c|c|c||c|c|c||c|c|c||c|c|c|}
\hline
$\alpha_{12} $ & $M,$ & $R_{s},$ & $\alpha $ & $M,$ & $R_{s},$ & $\alpha $ & $M,$ & $R_{s},$ & $\alpha_{12} $ & $M,$ & $R_{s},$\\
& $M_{\odot}$ & km & & $M_{\odot}$ & km & & $M_{\odot}$ & km & & $M_{\odot}$ & km\\
\hline
\multicolumn{3}{|c||}{$\rho_c = 10^7$ g/cm$^3$} & \multicolumn{3}{c|}{$\rho_c = 10^8$ g/cm$^3$} & \multicolumn{3}{|c||}{$\rho_c = 10^9$ g/cm$^3$} & \multicolumn{3}{c|}{$\rho_c = 10^{10}$ g/cm$^3$}\\
\hline
0 & 1.455 & 15530  & 0 & 1.453 & 7210 & 0 & 1.450 & 3340 & 0 & 1.443 & 1550  \\
\hline
1 & 1.455 & 15530 & 1 & 1.453 & 7210 & 1 & 1.448 & 3350 & 1 & 1.435 & 1560 \\
\hline
5 & 1.455 & 15530 & 5 & 1.451 & 7210 & 5 & 1.441 & 3360 & 5 & 1.403 & 1590 \\
\hline
10 & 1.454 & 15530 & 10 & 1.450 & 7220 & 10 & 1.432 & 3380 & 10 & 1.365 & 1630 \\
\hline
50 & 1.451 & 15560 & 50 & 1.434 & 7280 & 50 & 1.366 & 3520 & 50 & 1.142 & 2110 \\
\hline
\end{tabular}
\caption{\label{tab:singularity1}Masses and radii of stellar configurations at different values of central density and $\alpha$ for $n=3$.}
\end{center}
\end{table}
We calculated the relation between the mass and radius in the same
interval of the central density for $n=3/2$ as in the case of GR,
using the Jordan frame transformed quantities under study. Our
calculations indicate that only when the parameter $\alpha$ takes
values close to the upper limit of \cite{Naf}, significant
deviations from GR occur. In Fig. 3 we present the mass-radius
relation and the dependence of the mass from the central density
in comparison with the GR results. For a given mass, the radius of
the star decreases. Of course these results do not allow us having
hopes for pinpointing the underlying extended gravity theory from
real white dwarf observational data. The corresponding errors of
measurements for masses and radii have the same order as the
effect from $R^2$-gravity even for maximal values of the parameter
$\alpha$.

For $n=3$ and with the interval of the central densities being
between $10^7$ and $10^{10}$ g/cm$^3$, already for smaller values
of the parameter $\alpha$, the mass-radius diagram deviates from
the straight line corresponding to Chandrasekhar limit (see Fig.
4). For a given central density, the radius of star increases in
comparison with the radius in the case of GR. The parameters of
the stellar configurations for some central densities are given in
Table 1.

It is also interesting to investigate the mass and density
profiles in comparison with the GR results. From Fig. 5 one can
see that the effect from modified gravity is cumulative. Density
profiles in modified gravity and GR differ negligible but the
total mass inside a radius $r$, gradually decreases. {One note that results on Fig. 5 are given for $m$ as function of $r$ in Jordan frame.}
\begin{figure}
    \centering
    \includegraphics[scale=0.38]{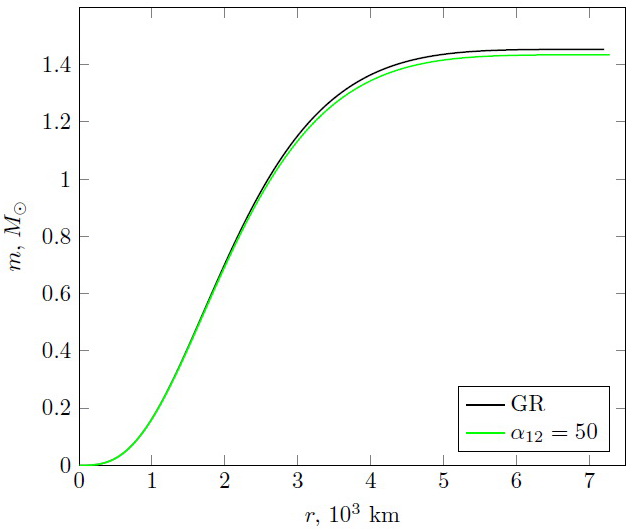}\includegraphics[scale=0.38]{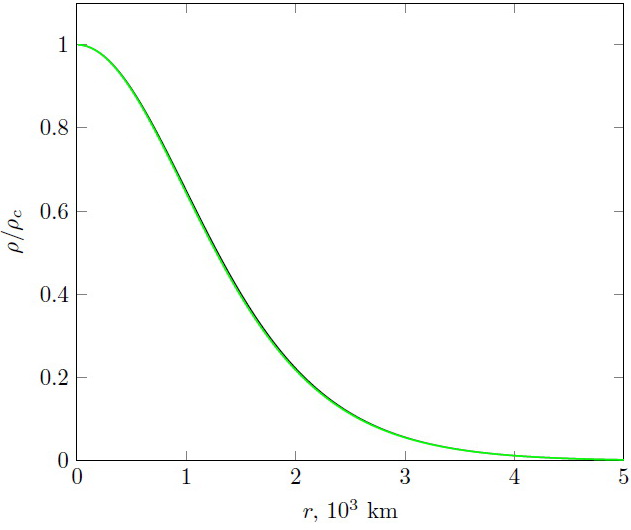}\\
    \includegraphics[scale=0.38]{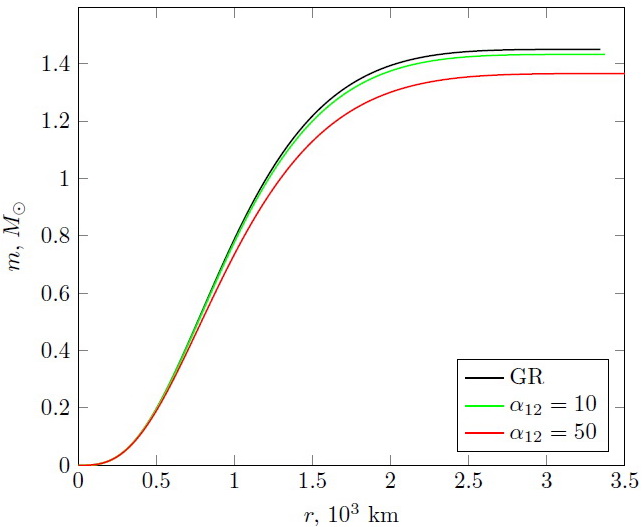}\includegraphics[scale=0.38]{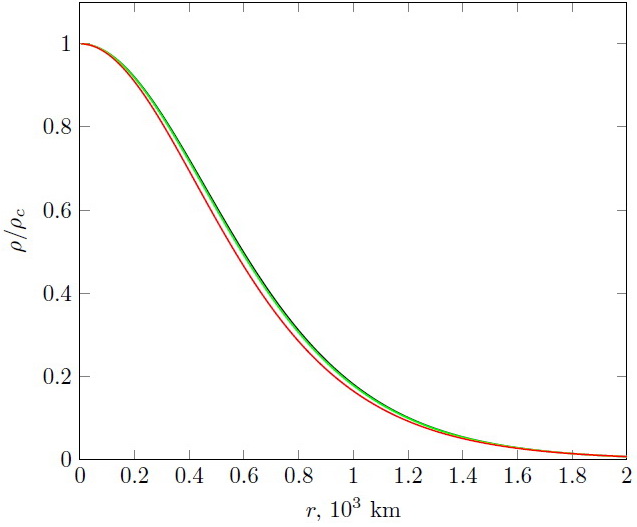}\\
    \includegraphics[scale=0.38]{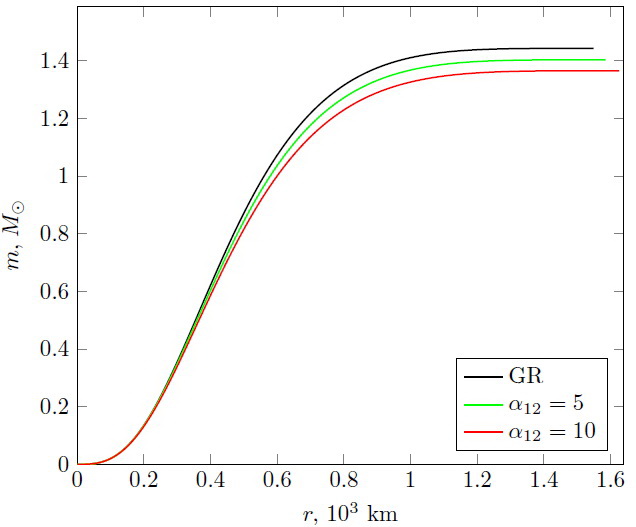}\includegraphics[scale=0.38]{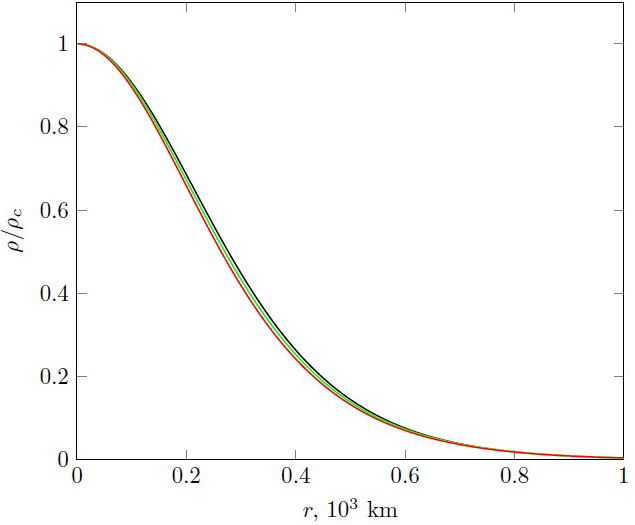}\\
    \caption{Profiles of the mass parameter in Jordan frame and density for $\rho_c=10^8$ g/cm$^3$
(upper panel), $\rho_c=10^9$ g/cm$^3$ (middle panel),
$\rho_c=10^{10}$ g/cm$^3$ (bottom panel) for some $\alpha$ in
units of $10^{12}$ cm$^2$ for $n=3$. For better visibility we cut
r-axis for coordinates where $\rho/\rho_{c}$ is negligible.}
    \label{fig:5}
\end{figure}
\begin{figure}
    \centering
    \includegraphics[scale=0.4]{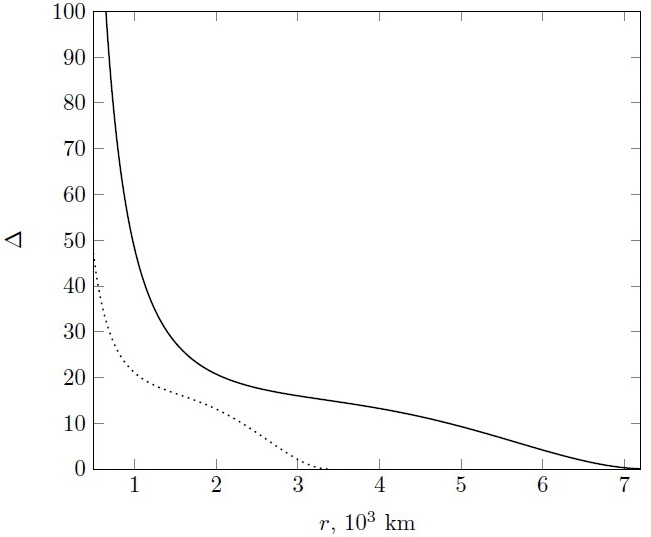}\includegraphics[scale=0.4]{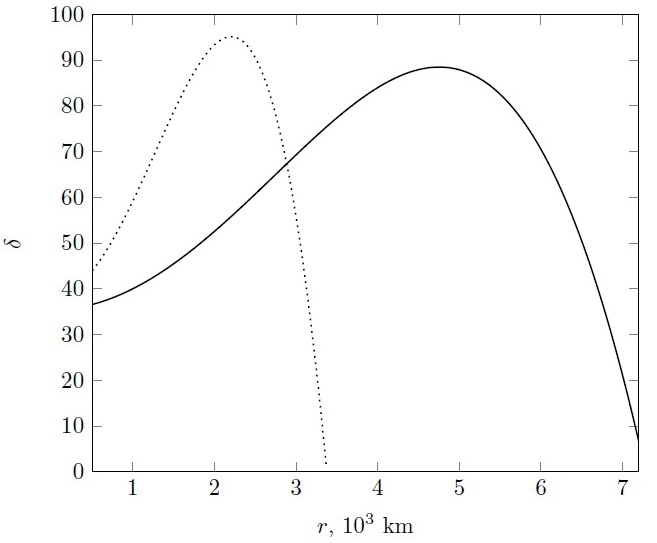}
    \caption{The relation between kinetic and potential terms
$\Delta=2{v}/{\phi'^{2}}$ (left) for $\rho_c=10^{8}$ g/cm$^3$,
$\alpha=5\times 10^{13}$ cm$^2$ (solid line) and $\rho_c=10^{9}$
g/cm$^3$, $\alpha=10^{13}$ cm$^2$ (dotted line). On the right
panel, the ratio $\delta=\frac{v}{\phi'/x}$ is shown.}
    \label{fig:6}
\end{figure}
The final step is the analysis of applicability of the
perturbative approach. For illustration we compared values of
potential $v$ and the term with the square of the derivative of
field $\phi$ for some parameters (Fig. 6). Only in the vicinity of
star's surface, the term $\phi'^{2}/2$ is comparable with
$v(\phi)$, but near the star surface, the effects of modified
gravity in any case are negligible. We also investigated relation
between $v(\phi)$ and term $\phi'/x$ in equation of scalar field.
Calculation therefore indicates that the simple approximation
(\ref{simple}) is valid for the scalar field $\phi$ and the values
of the parameters considered in this paper.

\section{Concluding Remarks}

For neutron stars in $R^2$-gravity, the maximal mass for
$\alpha\sim 10^{11}$ cm${^2}$ increases by a value around $\delta
M \sim 0.05-0.09$ $M_{\odot}$ ($\delta M$ depends from equation of
state for dense matter, see \cite{Alvaro}). The densities in
neutron star cores for maximal masses are  $\rho_c>3\times
10^{15}$ g/cm$^3$.

For relativistic polytropes in white dwarfs we obtain the same
order effects (but mass decreases) for central densities $\sim
10^{10}$ g/cm$^3$ (five orders of magnitude difference) although
for $\alpha$, we obtain two orders of magnitude larger ($\sim
10^{13}$ cm$^2$). This feature can be explained by the fact that
the characteristic size of white dwarfs is three orders greater in
comparison with neutron star. Additional terms in modified TOV
equations are smaller in comparison with NSs case, but due to the
long distance we obtain comparable effect.

It is interesting to note that our results concerning relativistic
polytropes are in concordance with calculations for NSs with
$M<1.5 M_{\odot}$. For densities $\rho_c<10^{15}$ g/cm$^3$, the
mass of the NSs in $R^2$ gravity decreases in comparison with the
GR case. This result can be explained by the stiffness of dense
matter EoS which decreases with density. One can conclude that
only for very large densities, the net effect of $R^2$ gravity
yields an increase of the gravitational mass. Although for more
precise calculations of white dwarfs parameters in $R^2$ gravity
one should consider more realistic EoS than polytropic ones, we
can expect that the main result of this work will remain the same:
the mass of the white dwarf decreases in comparison with GR.

Our results can be useful for establishing more stringent limits
on $R^2$ gravity. Now according to observations there are white
dwarf stars with masses larger than $1.3 M_{\odot}$. According to
\cite{Cilic} 25 such white dwarfs in vicinity of Sun System exist.
From realistic carbon monoxide core model the highest mass of
white dwarf is $1.334 M_{\odot}$ which corresponds to the
high-density limit $\rho = 10^9$ g/cm$^3$ (in GR). White dwarf
J1329 + 2549 is currently the most massive white dwarf known in
the solar neighborhood with well-constrained atmospheric
parameters and a mass of $1.351\pm 0.006 M_{\odot}$. These results
lead to conclusion that GR gives satisfactory picture for white
dwarfs parameters.

At first glance, this issue does not allow us to constrain
significantly $R^2$ gravity. But we can impose restrictions on the
parameter $\alpha$. The assumption that central density of
ultra-massive white dwarfs is around $10^9$ g/cm$^3$ and that
decreasing of mass due to the modification of gravity is the same
as in a case of relativistic polytrope for $\rho_c=10^9$ g/cm$^3$,
allow us to draw the following conclusion.: for $\alpha\sim
5\times 10^{13}$ cm$^2$ existence of white dwarfs with CO cores
and masses larger than $1.3 M_{\odot}$ would be questionable.
Unlike neutron stars, the EoS for matter in white dwarfs is
considered more well established and it is difficult to propose
realistic EoS which describes in GR white dwarfs with masses
$>1.33 M_{\odot}$. Therefore in principle we can estimate upper
limit on $\alpha$ as $\sim 10^{13}$ cm$^2$.

This is a quite interesting situation. From observational data for
white dwarf we obtained upper limit of an unknown modified gravity
parameter. On the contrary some indications to favor of possible
existence of supermassive NS with $M>2.2 M_{\odot}$ allow us to
estimate the lower limit of $\alpha$ although for dense nuclear
matter, the uncertainty in knowledge of the exact EoS is large
enough. Thus for testing modified gravity it is useful to consider
not only NSs but also less extreme objects like white dwarfs.

\section*{Acknowledgments}

This work was supported by MINECO (Spain), project
PID2019-104397GB-I00 (S.D.O). This work by S.D.O was also
partially supported by the program Unidad de Excelencia Maria de
Maeztu CEX2020-001058-M, Spain. This work was supported by
Ministry of Education and Science (Russia), project
075-02-2021-1748 (AVA).

\end{document}